\documentstyle[preprint,eqsecnum,aps,epsf]{revtex}
\begin{document}
\draft
\title{Second Generation Leptoquark Search\\
in $p\overline{p}$ Collisions at
$\sqrt{s}$ = 1.8 TeV\footnote{Submitted to Physical Review Letters} }

%
\author{
S.~Abachi,$^{12}$
B.~Abbott,$^{34}$
M.~Abolins,$^{23}$
B.S.~Acharya,$^{41}$
I.~Adam,$^{10}$
D.L.~Adams,$^{35}$
M.~Adams,$^{15}$
S.~Ahn,$^{12}$
H.~Aihara,$^{20}$
J.~Alitti,$^{37}$
% ...
\\
\vskip 0.50cm
\centerline{(D\O\ Collaboration)}
\vskip 0.50cm
}
\address{
\centerline{$^{1}$Universidad de los Andes, Bogot\'{a}, Colombia}
\centerline{$^{2}$University of Arizona, Tucson, Arizona 85721}
\centerline{$^{3}$Brookhaven National Laboratory, Upton, New York 11973}
\centerline{$^{4}$Brown University, Providence, Rhode Island 02912}
\centerline{$^{5}$University of California, Davis, California 95616}
\centerline{$^{6}$University of California, Irvine, California 92717}
\centerline{$^{7}$University of California, Riverside, California 92521}
\centerline{$^{8}$LAFEX, Centro Brasileiro de Pesquisas F{\'\i}sicas,
                  Rio de Janeiro, Brazil}
\centerline{$^{9}$CINVESTAV, Mexico City, Mexico}
\centerline{$^{10}$Columbia University, New York, New York 10027}
\centerline{$^{11}$Delhi University, Delhi, India 110007}
\centerline{$^{12}$Fermi National Accelerator Laboratory, Batavia,
                   Illinois 60510}
\centerline{$^{13}$Florida State University, Tallahassee, Florida 32306}
\centerline{$^{14}$University of Hawaii, Honolulu, Hawaii 96822}
\centerline{$^{15}$University of Illinois at Chicago, Chicago, Illinois 60607}
\centerline{$^{16}$Indiana University, Bloomington, Indiana 47405}
\centerline{$^{17}$Iowa State University, Ames, Iowa 50011}
\centerline{$^{18}$Korea University, Seoul, Korea}
\centerline{$^{19}$Kyungsung University, Pusan, Korea}
\centerline{$^{20}$Lawrence Berkeley Laboratory and University of California,
                   Berkeley, California 94720}
\centerline{$^{21}$University of Maryland, College Park, Maryland 20742}
\centerline{$^{22}$University of Michigan, Ann Arbor, Michigan 48109}
\centerline{$^{23}$Michigan State University, East Lansing, Michigan 48824}
\centerline{$^{24}$Moscow State University, Moscow, Russia}
\centerline{$^{25}$University of Nebraska, Lincoln, Nebraska 68588}
\centerline{$^{26}$New York University, New York, New York 10003}
\centerline{$^{27}$Northeastern University, Boston, Massachusetts 02115}
\centerline{$^{28}$Northern Illinois University, DeKalb, Illinois 60115}
\centerline{$^{29}$Northwestern University, Evanston, Illinois 60208}
\centerline{$^{30}$University of Notre Dame, Notre Dame, Indiana 46556}
\centerline{$^{31}$University of Oklahoma, Norman, Oklahoma 73019}
\centerline{$^{32}$University of Panjab, Chandigarh 16-00-14, India}
\centerline{$^{33}$Institute for High Energy Physics, 142-284 Protvino, Russia}
\centerline{$^{34}$Purdue University, West Lafayette, Indiana 47907}
\centerline{$^{35}$Rice University, Houston, Texas 77251}
\centerline{$^{36}$University of Rochester, Rochester, New York 14627}
\centerline{$^{37}$CEA, DAPNIA/Service de Physique des Particules, CE-SACLAY,
                   France}
\centerline{$^{38}$Seoul National University, Seoul, Korea}
\centerline{$^{39}$State University of New York, Stony Brook, New York 11794}
\centerline{$^{40}$SSC Laboratory, Dallas, Texas 75237}
\centerline{$^{41}$Tata Institute of Fundamental Research,
                   Colaba, Bombay 400005, India}
\centerline{$^{42}$University of Texas, Arlington, Texas 76019}
\centerline{$^{43}$Texas A\&M University, College Station, Texas 77843}
}
\date{\today}
\maketitle
\begin{abstract}
We report on a search for second generation leptoquarks
with the D\O\ detector at the Fermilab Tevatron $p\overline{p}$
collider at $\sqrt{s}$ = 1.8 TeV.
This search is based on 12.7 pb$^{-1}$ of data.
Second generation leptoquarks are assumed to be produced in
pairs and to decay into a muon and quark with
branching ratio $\beta$ or to neutrino and quark with branching ratio
$(1-\beta)$.
We obtain cross section times branching ratio limits as a
function of leptoquark mass
and set a lower limit on the leptoquark mass of 111 GeV/c$^{2}$
for $\beta = 1 $ and
89 GeV/c$^{2}$ for $\beta = 0.5 $ at the 95\%\ confidence level.

\end{abstract}
\pacs{PACS numbers: 14.80.-j, 13.85.Rm, 12.10.-g}

 Leptoquarks are bosons predicted~\cite{theory} in many extensions of the
Standard Model (SM). They carry both lepton and color quantum
numbers and couple to leptons and quarks.
In order to satisfy
experimental constraints
on flavor changing neutral currents and rare pion decays,
leptoquarks are required to be left or
right handed and couple to only one
generation of leptons and quarks~\cite{hewett}.
These constraints are required unless leptoquarks are considerably
more massive~\cite{buch}
than particles the current Tevatron run can produce.

       This paper reports the results of a search for second
generation scalar leptoquarks.
We assume leptoquarks are produced in pairs by QCD processes.
At the Tevatron these QCD processes
dominate other production mechanisms which depend on the
leptoquark--lepton--quark coupling.
Second generation leptoquarks are assumed to decay with branching
ratio $\beta$ to a
muon and quark and with branching ratio $(1-\beta)$ to a
neutrino and quark.
There are three decay signatures for pair produced second generation
leptoquarks:
two muons plus at least two jets, one muon plus missing transverse
energy (\mbox{$E\kern-0.57em\raise0.19ex\hbox{/}_{T}$}) and at least two jets,
or \mbox{$E\kern-0.57em\raise0.19ex\hbox{/}_{T}$}\ plus two or more jets.
The muons, \mbox{$E\kern-0.57em\raise0.19ex\hbox{/}_{T}$}, and jets are
expected to
be well separated which distinguishes pair produced leptoquarks from
$c$-quark and $b$-quark production, and
leptoquarks are distinct from $W$ and $Z$ boson backgrounds
due the presence of at least two
energetic jets.
This report gives results for limits on cross section times
$\beta^{2}$ for the dimuon signature and cross section times
$2\beta(1-\beta)$ for the single muon signature. These cross
section limits are used to set limits on the second generation leptoquark mass.
The data used for this analysis were taken during the Tevatron run between
August 1992 and May 1993 and represent an integrated
luminosity of 12.7 pb$^{-1}$.

Previous limits from LEP experiments exclude leptoquark masses
below 45 GeV/c$^{2}$~\cite{lep}. Results from \mbox{D\O } and CDF
on first generation scalar leptoquarks have been published~\cite{first}.
The \mbox{D\O } limits for first generation leptoquarks are 133 GeV/c$^{2}$ and
120 GeV/c$^{2}$ for $\beta$ = 1.0 and 0.5, and the CDF limits are
113 GeV/c$^{2}$ and 80 GeV/c$^{2}$ for $\beta$ = 1.0 and 0.5.
Experiments at HERA~\cite{hera}
give limits on the mass of first generation leptoquarks where their limits
depend on the
unknown but constrained~\cite{eelq}
leptoquark--electron--quark coupling which they have generally assumed to
be at the strength of the electro--weak coupling.

The \mbox{D\O } detector, described in detail elsewhere~\cite{detect},
is composed of three major systems:
an inner detector (without a magnetic field) for tracking charged particles
within a pseudorapidity range $\mid \eta \mid < 3.5$, a calorimeter
for measuring electromagnetic and hadronic showers within the range
$\mid \eta \mid < 4.0$, and a muon spectrometer covering the range
$\mid \eta \mid < 3.3$. The calorimeter has fine segmentation in both $\eta$
and
azimuth, $\phi$, and measures
electrons with a resolution of 15\%/$\sqrt{E}$ and hadrons with a
resolution of about 50\%/$\sqrt{E}$.
Muons are identified and their momentum, $p$, measured with three layers of
proportional drift tubes,
one before (coming from the interaction region) and two after the
magnetized iron
toroids. The muon momentum resolution is
$\sigma (1/p) = 0.18(p-2)/p^{2} \oplus 0.008$
(with $p$ in GeV/c).

 Muons are required to have an impact parameter consistent
with coming from the interaction region and to have $\mid \eta \mid < 1.7$.
Cosmic
ray muons are removed by requiring that there are no tracks or pattern of hits
back-to-back in $\eta$ and $\phi$. Muons are required to be
well isolated to reduce backgrounds from heavy quark production by first
requiring that there be no jet with transverse energy (\mbox{$E_{T}$})
greater than 20 GeV
within an angle of 0.7 radians of the muon and, secondly, that the expected
deposited energy along the muon track in the calorimeter is not more than
about three times that expected from a minimum ionizing particle (about
3 GeV from a MIP for the \mbox{D\O } calorimeter).
A further set of requirements is defined for
high quality muon identification.  This includes requiring energy deposition in
the calorimeter consistent with a minimum ionizing particle and
a matching track in the tracking chamber. The muon must hit all three
layers of the muon system and have timing consistent with originating
from the beam crossing. Finally, the muon must traverse a minimum field
integral of 1.83 T$\cdot$m of the toroid magnet.

Jets are measured in the calorimeter. They are defined
by a cone algorithm with ${\cal R} = \sqrt{(\Delta \eta)^{2}
+(\Delta \phi)^2}$ = 0.7. Jets are corrected for
calorimeter response, underlying event, and out-of-cone leakage effects.
These corrections amount to about 25\% and vary with jet energy and
$\eta$. Jets are accepted within $\mid \eta \mid < 3.5$.

The \mbox{$E\kern-0.57em\raise0.19ex\hbox{/}_{T}$},
representing the transverse energy carried by the neutrino in the single
muon signature, is required to be isolated from the jets in $\phi$ by 0.3
radians. Also, the magnitude of the angular separation in $\phi$ of the muon
and \mbox{$E\kern-0.57em\raise0.19ex\hbox{/}_{T}$}\ cannot be greater than
$\pi - 0.2$. These cuts ensure that the
\mbox{$E\kern-0.57em\raise0.19ex\hbox{/}_{T}$}\ is not an artifact of either
fluctuations in the jet energy or the muon resolution.

\begin{figure}
\epsfxsize=7in
\epsfysize=6in
\epsffile{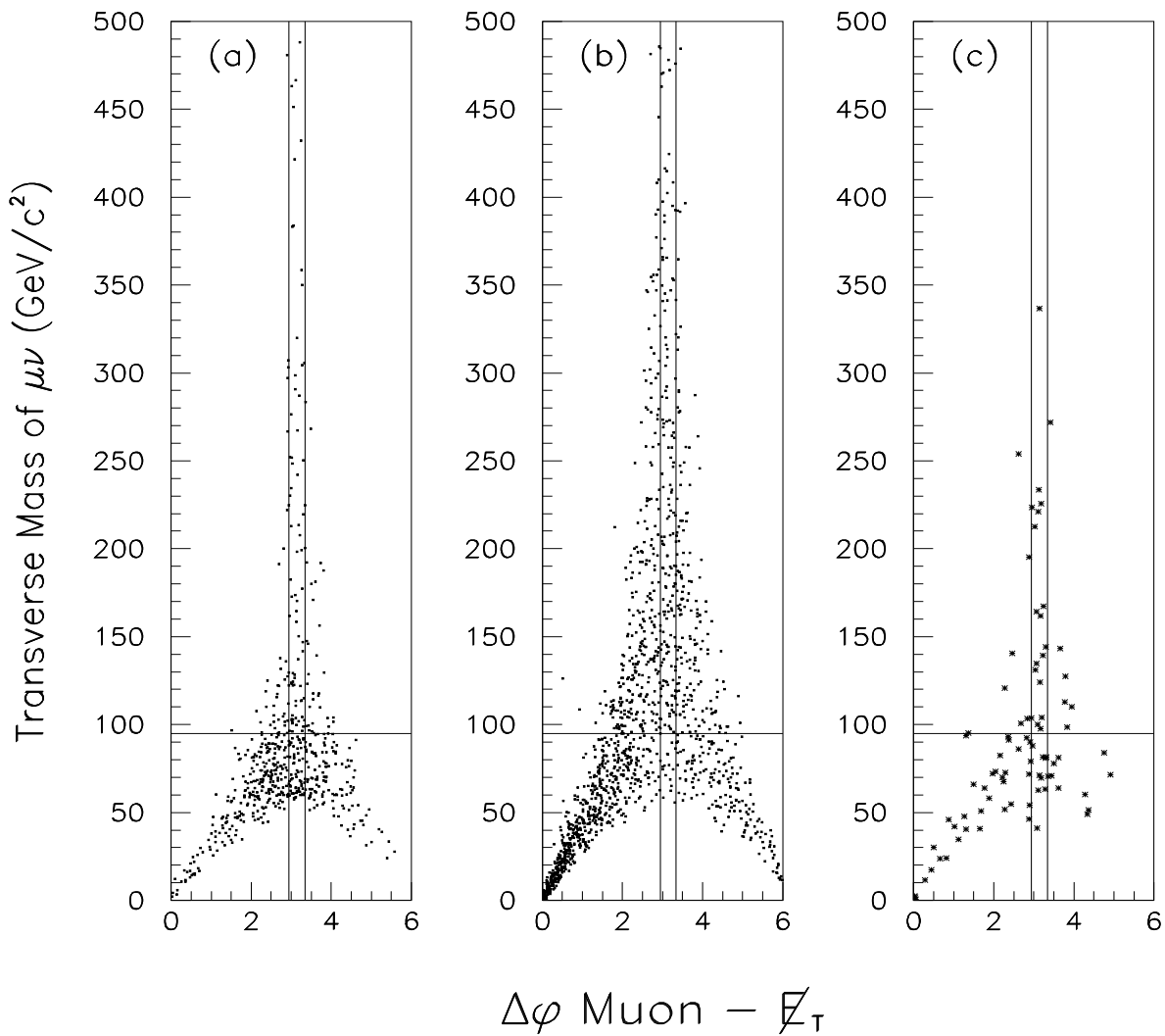}
\caption{$M_{T}$ versus the absolute difference in $\phi$ between the muon
and \mbox{$E\kern-0.57em\raise0.19ex\hbox{/}_{T}$}\ for
(a) a W$\rightarrow \mu \nu$ plus jets Monte Carlo sample,
(b) a 100 GeV/c$^{2}$ mass second generation leptoquark
Monte Carlo sample, and (c) for a data sample obtained by
the single muon signature selection with 10 GeV jets.
The horizontal and vertical lines show cuts used in the
analysis (see text for details) .  The number of events in
(a), (b), and (c) are not normalized to the same integrated luminosity.}
   \label{fig:tmphi}
\end{figure}

For the search in the single muon channel, the events in the
data sample are
required to pass a trigger with a muon transverse momentum (\mbox{$p_{T}$})
threshold of 8 GeV/c and a jet \mbox{$E_{T}\: \,$}threshold of 15 GeV.
Offline, one high quality muon with \mbox{$p_{T}\: \,$}$>$
20 GeV/c and $\mid \eta \mid < 1.0$ is required.
The \mbox{$E\kern-0.57em\raise0.19ex\hbox{/}_{T}$}\ is required to be
greater than 25 GeV. Having applied these kinematic cuts
and requiring the two leading jets to have \mbox{$E_{T}\: \,$}$>$ 10 GeV,
Fig.~\ref{fig:tmphi} shows the transverse mass ($M_{T}$) of
(\mbox{$E\kern-0.57em\raise0.19ex\hbox{/}_{T}$}\ + $\mu$) versus the
absolute difference between
the $\phi$ of the \mbox{$E\kern-0.57em\raise0.19ex\hbox{/}_{T}$}\ and the muon
for three event samples:
$W$ boson plus jets Monte Carlo, single muon leptoquark Monte Carlo
with a mass of 100 GeV/c$^{2}$, and the data.
The two vertical lines indicate the region removed by the
muon-\mbox{$E\kern-0.57em\raise0.19ex\hbox{/}_{T}$}\ back-to-back $\phi$ cut,
and the horizontal line indicates a $M_{T}$ cut where we require that
leptoquark
candidates have $M_{T}$ greater than 95 GeV/c$^{2}$.
Since we expect two high \mbox{$E_{T}\: \,$}jets from leptoquarks, the
\mbox{$E_{T}\: \,$}requirement on the jets is raised to 25 GeV.
These jets are also required to have an electromagnetic energy fraction
greater than 0.2 to reduce the backgrounds from large jet energy fluctuations
or calorimeter noise which may not be correlated with the
\mbox{$E\kern-0.57em\raise0.19ex\hbox{/}_{T}$}.
No candidates remain after the jet \mbox{$E_{T}\: \,$}cut.

For the single muon signature, the expected backgrounds
come from $W$ boson plus jets production, leptonic decays of $b\bar{b}$ pairs,
Drell-Yan dimuon plus jets production
where one muon is missing,
and the decay of $W$ and $Z$ bosons into heavy quarks with semileptonic decays.
The number of expected background events for the single muon signature
is given in Table~\ref{tab:backmn} for a few values of the
$M_{T}$ cut. The background estimates reasonably account for the data.

For the dimuon signature selection, the candidate events are required to
pass the same trigger
as the single muon events.
In the offline selection for
the dimuon sample, both muons are required to be isolated and to
have a \mbox{$p_{T}\: \,$}greater than 25 GeV/c. At least one muon is
required to pass the
high quality cuts,
and at least one  is required to have $\mid \eta \mid < 1.0$.
These cuts leave 15 events
in the dimuon sample. For the leptoquark signature,
we require that our candidate events have at least two
jets with \mbox{$E_{T}\: \,$}greater than 25 GeV. This jet cut
significantly reduces the
Drell-Yan sources of background for this signature. With this last cut
no candidate events are left.

The main sources of background for the dimuon signature are Drell-Yan
dimuons plus jets and
leptonic decays of $b\bar{b}$ pairs.
Background estimates are made for different kinematic cuts.
In Table~\ref{tab:backmm}, the background estimates for 15, 20, and 25 GeV
cuts on the muons and jets in the dimuon sample are shown with the actual
number of events seen.
The estimated backgrounds reasonably account for the data.

The efficiencies of the cuts used in the selection for the two signatures
are determined from a study of Monte Carlo generated and collider events.
The geometric acceptance and kinematic efficiency are taken from
leptoquark signal
Monte Carlo generated by {\small ISAJET}~\cite{isajet} and processed
with
a \mbox{D\O } version of the {\small GEANT}~\cite{geant} detector simulator,
a simulation
of the \mbox{D\O } trigger, and \mbox{D\O }'s standard reconstruction program.
The efficiencies
for muon identification are determined
from the data; they amount to about 21\% for the dimuon selection and 32\%
for the single muon
selection. The trigger efficiencies calculated from a study of both the data
and the Monte Carlo are $86.2\pm0.82$\% for the dimuon signature and
$66.6\pm 1.6$\% for the single muon signature.
For the dimuon signature, the total efficiency ranges from 0.35\%
to 8.7\% for leptoquark masses  between 45 and 200 GeV/c$^{2}$. For the
single muon signature the total efficiency ranges from 0.14\% to 5.12\%
for the same mass range. The relative
uncertainty
on the total efficiency is 20\% for the dimuon signature and 10\% for the
single muon signature. This uncertainty on the efficiency is dominated by the
statistics of the $Z\rightarrow \mu^{+} \mu^{-}$ data sample used to calculate
the efficiency of the muon quality cuts. The systematic uncertainties
vary from 27\% to 9\% for the dimuon channel for leptoquark masses
ranging from 45 to 200 GeV/c$^{2}$.
These systematic uncertainties arise
from a 10\% jet energy scale uncertainty and a 10\% and 25\% uncertainty in
the first and second terms of the muon \mbox{$p_{T}\: \,$}resolution.
For the single muon signature the systematic
uncertainties vary from 16\% to 12\% for the same mass range.
For both signatures the dominant systematic effect comes from the
uncertainty in the muon \mbox{$p_{T}\: \,$}resolution.

\begin{figure}
\epsfxsize=6in
\epsfysize=6in
\epsffile{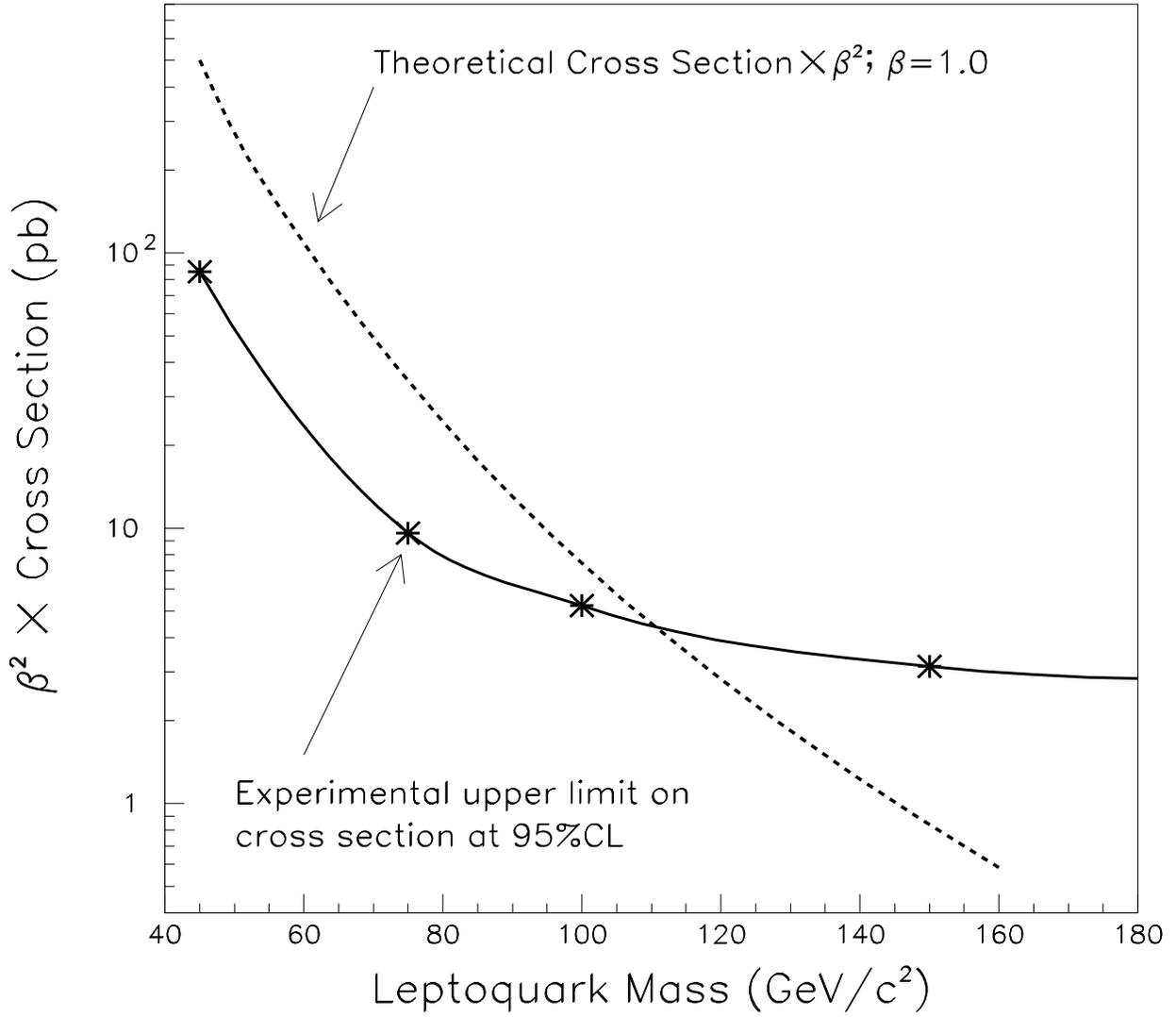}
\caption{The 95\% CL upper limit obtained by \mbox{D\O }
on the cross section times $\beta ^{2}$
for the dimuon signature,
as a function of the leptoquark mass.
Also shown is the {\protect\small ISAJET} prediction times $\beta ^{2}$
for $\beta$=1.0}
   \label{fig:one}
\end{figure}

\begin{figure}
\epsfxsize=6in
\epsfysize=6in
\epsffile{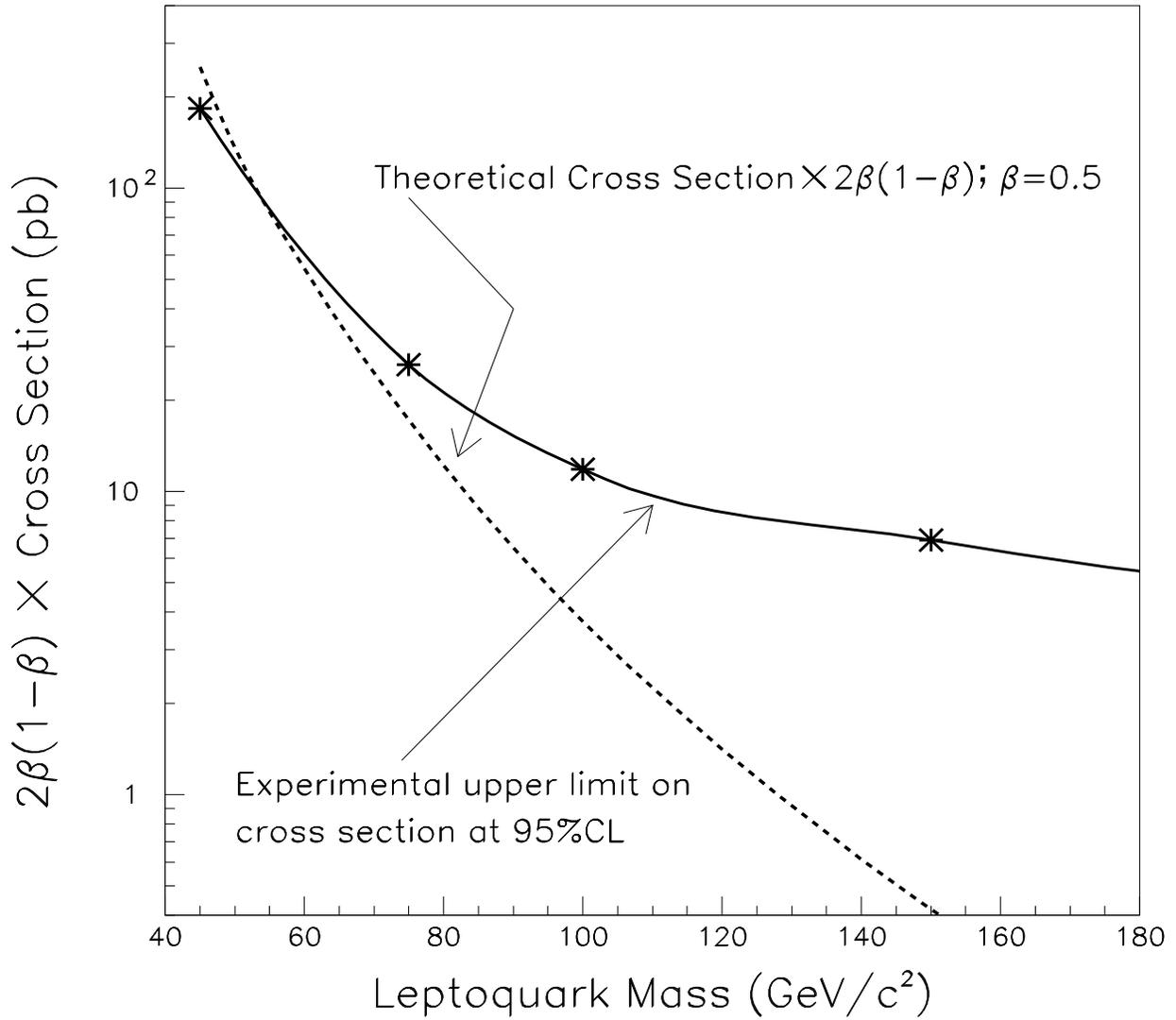}
\caption{The 95\% CL upper limit on the cross section times 2$\beta (1-\beta)$
for the single muon signature,
as a function of the leptoquark mass.
Also shown is the  {\protect\small ISAJET} prediction times
2$\beta (1-\beta)$ for $\beta$=0.5}
   \label{fig:two}
\end{figure}

The 95\% confidence level (CL) limit on cross section times branching ratio
factor $\beta^{2}$ as a function of leptoquark mass for the
dimuon signature is given in Fig.~\ref{fig:one}. This cross section
limit takes into account the uncertainty in the integrated luminosity times
acceptance~\cite{cousins}
taken as the sum in quadrature of the above uncertainties including a
5.4\% systematic uncertainty in the integrated luminosity.
Also plotted in
Fig.~\ref{fig:one} is $\beta ^{2}$ times the theoretical cross section based on
{\small ISAJET}\cite{cs}
using the Morfin and Tung leading order (MT-LO) parton distribution functions
(pdf)~\cite{pdf}
for $\beta = 1$. The intersection of these two curves at a leptoquark
mass of 111 GeV/c$^{2}$ gives the 95\% CL lower limit on the
mass of a second generation leptoquark for $\beta = 1$.
The 95\% CL lower limit on cross section times branching ratio
factor $2\beta (1-\beta)$ as a function of leptoquark mass for the
single muon signature is given in Fig.~\ref{fig:two}.
For $\beta = 0.5$, the single muon limit is 54 GeV/c$^{2}$.

\begin{figure}
\epsfxsize=6in
\epsfysize=6in
\epsffile{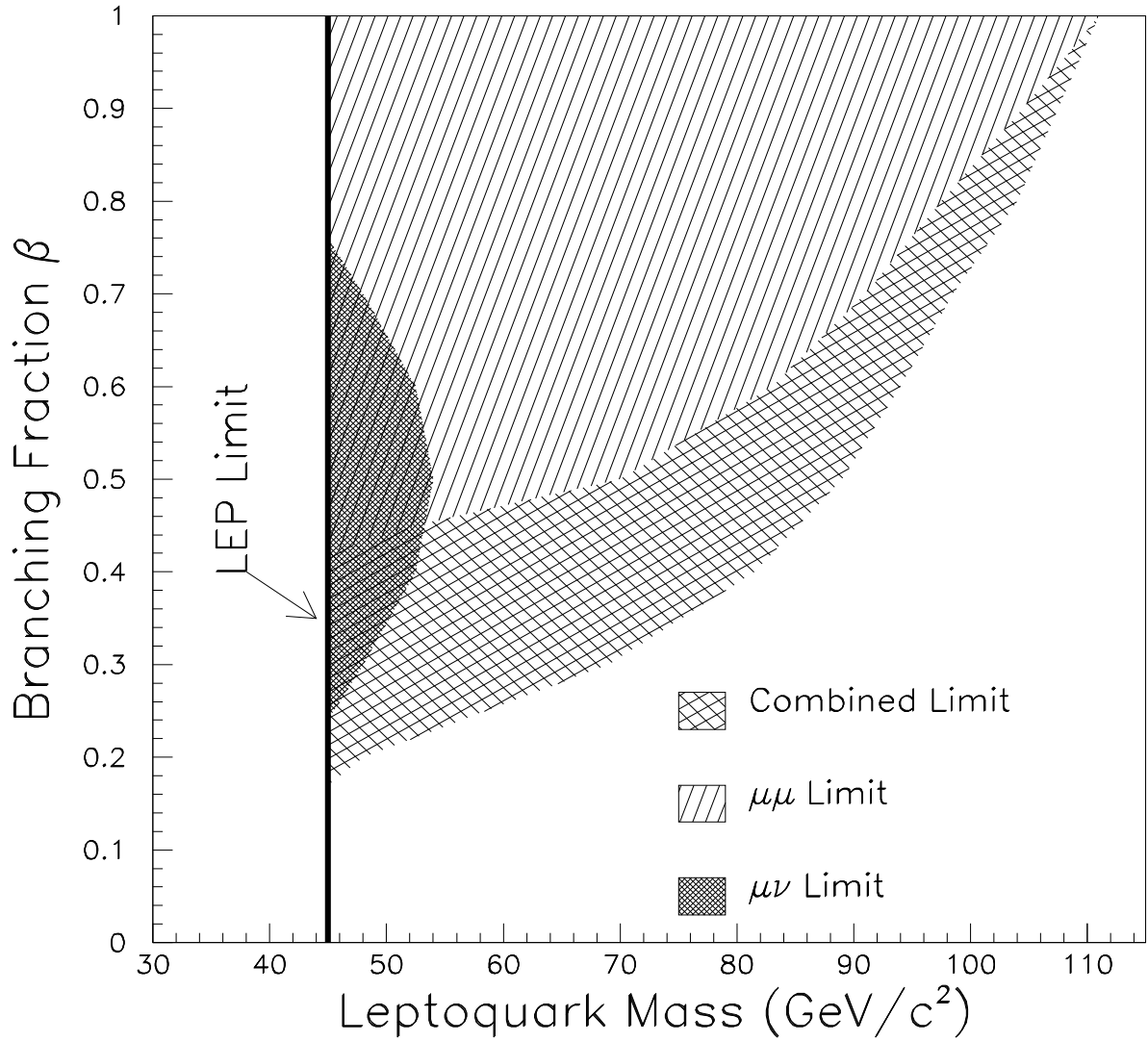}
  \caption{The 95\% CL excluded regions for the dimuon, single muon,
and combined signatures. }
   \label{fig:three}
\end{figure}

In Fig.~\ref{fig:three} we show the $\beta$ versus mass excluded region for the
dimuon signature as the area covered by the diagonal lines.
The area covered by the solid shading is the region excluded
for the single muon signature.
By combining the acceptance for the
single muon and dimuon signatures, we exclude the additional region
indicated by the cross hatched area.
The combined mass limit for $\beta=0.5$ is 89 GeV/c$^{2}$. The LEP limit of 45
GeV/c$^{2}$ is also given in Fig.~\ref{fig:three}. Our limit extends
to a branching fraction of $\beta=0.17$ at the LEP mass limit.
CDF~\cite{cdf}, based on the dimuon channel only, has also set limits on the
mass of second generation leptoquarks of 131 and 96 GeV/c$^{2}$ for
$\beta=1.0$ and $0.5$.
The mass limits depend somewhat on the choice of pdf,
momentum transfer scale (we have assumed
$Q^{2}=\hat{s}$ for this analysis), and higher order effects~\cite{mont}
which we have chosen not to include.
Using the same theoretical cross section and choice of pdf
(CTEQ2pM)
as quoted by CDF in Ref.~\cite{cdf}, our
combined mass
limits become 119 and 97 GeV/c$^{2}$ for $\beta = 1.0$ and $0.5$.
Using this theoretical cross section we can exclude, compared
to CDF, additional $\beta$ vs mass space starting at $\beta = 0.5$
and extending down to $\beta = 0.17$ at the LEP limit.

In conclusion we observe no events from second generation leptoquarks.
We have set limits on the mass as a function of $\beta$ for the pair
production of second generation leptoquarks where the cross
section for their production is independent of the coupling strength of
the leptoquark to a
second generation lepton and quark.

We thank the Fermilab Accelerator, Computing, and Research Divisions, and
the support staffs at the collaborating institutions for their contributions
to the success of this work.   We also acknowledge the support of the
U.S. Department of Energy,
the U.S. National Science Foundation,
the Commissariat \`a L'Energie Atomique in France,
the Ministry for Atomic Energy and the Ministry of Science and
Technology Policy in Russia,
CNPq in Brazil,
the Departments of Atomic Energy and Science and Education in India,
Colciencias in Colombia, CONACyT in Mexico,
the Ministry of Education, Research Foundation and KOSEF in Korea
and the A.P. Sloan Foundation.

\begin{table}
\caption{The number of single muon events is given in this table as a
function of $M_{T}$ (GeV/c$^{2}$) cut. All other cuts are kept the same as
given in the text. Also
given is the number of events expected from $W\rightarrow \mu \nu$ plus
jets, $b\bar{b}$, $Z\rightarrow \mu^{+}\mu^{-}$ plus jets where one
muon is missing,
$W\rightarrow$ $c$ $s$ $\rightarrow \mu$ plus jets and the total
expected backgrounds. Note that the uncertainty in the $Z$ background
is 100\% .}
\label{tab:backmn}
\begin{tabular}{|c|c|c|c|c||c|c|}
$M_{T}$  & $W$ & $b\bar{b}$ & $Z$ & $W\rightarrow$ $c$ $s$
& total bgd. & \# events\\
\hline
95 & 1.4$\pm$0.3 & 0.5$\pm$0.2     & 0.10& 0.37$\pm$0.37 & 2.4$\pm$1.0 & 0 \\
\hline
85 & 2.2$\pm$0.5 & 1.0$\pm$0.3    &0.13  & 0.37$\pm$0.37 & 3.7$\pm$1.3 & 3 \\
\hline
75 & 3.3$\pm$0.8 & 1.4$\pm$0.5    &0.17 & 0.74$\pm$0.54 & 5.6$\pm$2.0 & 5\\
\end{tabular}
\end{table}

\begin{table}
\caption{Estimates of background contributions to the
dimuon sample from Drell-Yan $\mu^{+}\mu^{-}$ with jets
(including $Z \to \mu^{+}\mu^{-}$)
and leptonic $b\bar{b}$ decays for the indicated threshold cuts on both the
two muons and two jets. Also
given is the number of dimuon plus jets events surviving these
threshold cuts.}
\label{tab:backmm}
\begin{tabular}{|c|c|c|c|}
$\mu$ \mbox{$p_{T}$}, jet \mbox{$E_{T}\: \,$}(GeV) &  Drell-Yan
& $b\bar{b}$  &\# events    \\
\hline
25  & 1.8$\pm$0.7  & 0.05$\pm$0.02  & 0  \\
\hline
20  & 3.4$\pm$1.0  & 0.23$\pm$0.11 & 3  \\
\hline
15 & 9.9$\pm$2.1  & 0.77$\pm$0.38  & 12   \\
\end{tabular}
\end{table}


\end{document}